\documentclass{ws-p8-50x6-00}
\usepackage{graphicx}

\def\bbuildrel#1_#2^#3{\mathrel{\mathop{\kern 0pt#1}\limits_{#2}^{#3}}}
\def\slash#1{\setbox0=\hbox{$#1$}#1\hskip-\wd0\dimen0=5pt\advance
       \dimen0 by-\ht0\advance\dimen0 by\dp0\lower0.5\dimen0\hbox
         to\wd0{\hss\sl/\/\hss}}

\newcommand{\f}{\frac}
\newcommand{\als}{\alpha_s}

\begin{document}
\title{Theory of radiative $B$ decays}
\author{MIKO{\L}AJ MISIAK}
\address{Theory Division, CERN, CH-1211 Geneva 23, Switzerland \\and\\
Institute of Theoretical Physics, Warsaw University, 00-681 Warsaw, Poland}
\maketitle \abstracts{Theory of charmless radiative B decays is
  reviewed.  Existence of uncontrolled non-perturbative effects in the
  inclusive rate at ${\cal O}(\als)$ is reminded.}

\begin{figure}[h]
\begin{center}
\includegraphics[width=10cm,angle=0]{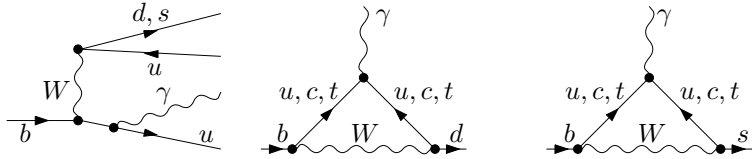}
\caption{Examples of Feynman diagrams contributing to $\bar{B} \to X_{\rm no~charm} \gamma$.}
\label{types}
\end{center}
\vspace{-32mm}
\hspace*{25mm} $d,s$ \hspace{25mm} $\gamma$ \hspace{33mm} $\gamma$\\[3mm]
\hspace*{27mm} $u$\\[-1mm]
\hspace*{13mm} $W$ \hspace{9mm} $\gamma$
\hspace{12mm} $u,c,t$ \hspace{5mm} $u,c,t$ 
\hspace{12.5mm} $u,c,t$ \hspace{5mm} $u,c,t$\\[1.5mm]
\hspace*{1cm} $b$ \hspace{2cm} $u$ 
\hspace{6mm}  $b$ \hspace{6mm} $W$ \hspace{7mm} $d$ 
\hspace{11mm} $b$ \hspace{6mm} $W$ \hspace{7mm} $s$\\[4mm]
\end{figure}

Charmless radiative $\bar{B}$ decays are the decays $\bar{B} \to
X_{\rm no~charm} \gamma$, where $X_{\rm no~charm}$ is either a
particular hadronic state for an exclusive decay, or just any
charmless hadronic state in the inclusive case. Such decays are
generated by tree-level $b \to u$ transitions with photon radiation,
loop-mediated $b \to d$ transitions, and loop-mediated $b \to s$
transitions. Examples of diagrams contributing to each of these three
types of transitions are presented in Fig.~\ref{types}. The
loop-mediated transitions are known to be very sensitive to new
physics, e.g. to existence of SUSY particles with masses below 1~TeV.
In the Standard Model, the $b \to u$ and $b \to d$ transitions are
CKM-suppressed with respect to the $b \to s$ ones. The relative
suppression factors are $|V_{ub}/V_{ts}|^2 \simeq 1\%$ and
$|V_{td}/V_{ts}|^2 \in [2.5\%,5\%]$, respectively.  Therefore, to a
good approximation, the loop-mediated $b \to s$ transitions saturate
the inclusive $\bar{B} \to X_{\rm no~charm} \gamma$ branching ratio.
This branching ratio is measured by CLEO\cite{CLEO99a} and
ALEPH\cite{ALEPH98} at the level of around $3\times10^{-4}$, after the
contribution from intermediate $\psi$ states is subtracted.
\bea 
\sum_{X_s} BR[\bar{B} \to X_s \gamma]_{\rm loop~mediated}
&\simeq& \sum_{X_{\rm no~charm}} BR[\bar{B} \to X_{\rm no~charm} \gamma]
\nonumber\\ && \hspace{-3cm}
\simeq (3.15 \pm 0.35 \pm 0.32 \pm 0.26) \times 10^{-4} 
                        ~~~~~{\rm \bf (CLEO)}^1
\nonumber\\ && \hspace{-2cm} \label{first}
+      BR[\bar{B} \to X^{(1)}_{\rm no~charm} \psi  ]
\times BR[ \psi   \to X^{(2)}_{\rm no~charm} \gamma].
\eea
The latter term in the above equation is the very contribution from
intermediate $\psi$. A lower bound on its numerical size can be found
by summing up the exclusive branching ratios of charmless radiative
$\psi$ decays listed in the Particle Data Book\cite{PData98} (which
give together around 4\%) and multiplying them by $BR[\bar{B} \to X
\psi]$ that is close to 1\%. It follows that the intermediate $\psi$
contribution to $BR[\bar{B} \to X_s \gamma]$ is not smaller than $4
\times 10^{-4}$, i.e. it is larger than the remainder of this
branching ratio.

The CLEO result is interpreted here as the one with subtracted
intermediate $\psi$ contribution, even though no subtraction has been
actually made in the measurement performed for high-energy
photons.\cite{TS99} However, the extrapolation to lower photon
energies has been done with use of a theoretical model that did not
include intermediate $\psi$. Photons originating from the intermediate
$\psi$ are expected to be softer than most of the other photons in
$\bar{B} \to X_s \gamma$. A quantitative estimate of their softness
(missing at present) is necessary to completely clarify this point.

The intermediate $\psi$ contribution had to be included in
Eq.~(\ref{first}), because each $X_{\rm no~charm}$ is assumed to be a
QCD-eigenstate, while $\psi$ is not stable in QCD.  In other words,
the diagrams in Fig.~\ref{types}, when dressed by an appropriate
number of gluons, give a contribution to the intermediate $\psi$
channel as well. Thus, a separation of the intermediate $\psi$
contribution, which may be straightforward on the experimental side,
has to be thought about on the theoretical side,
too.\cite{LRW97,LLMW99} We shall come back to this point later.

\begin{figure}[t]
\begin{center}
\includegraphics[width=4cm,angle=0]{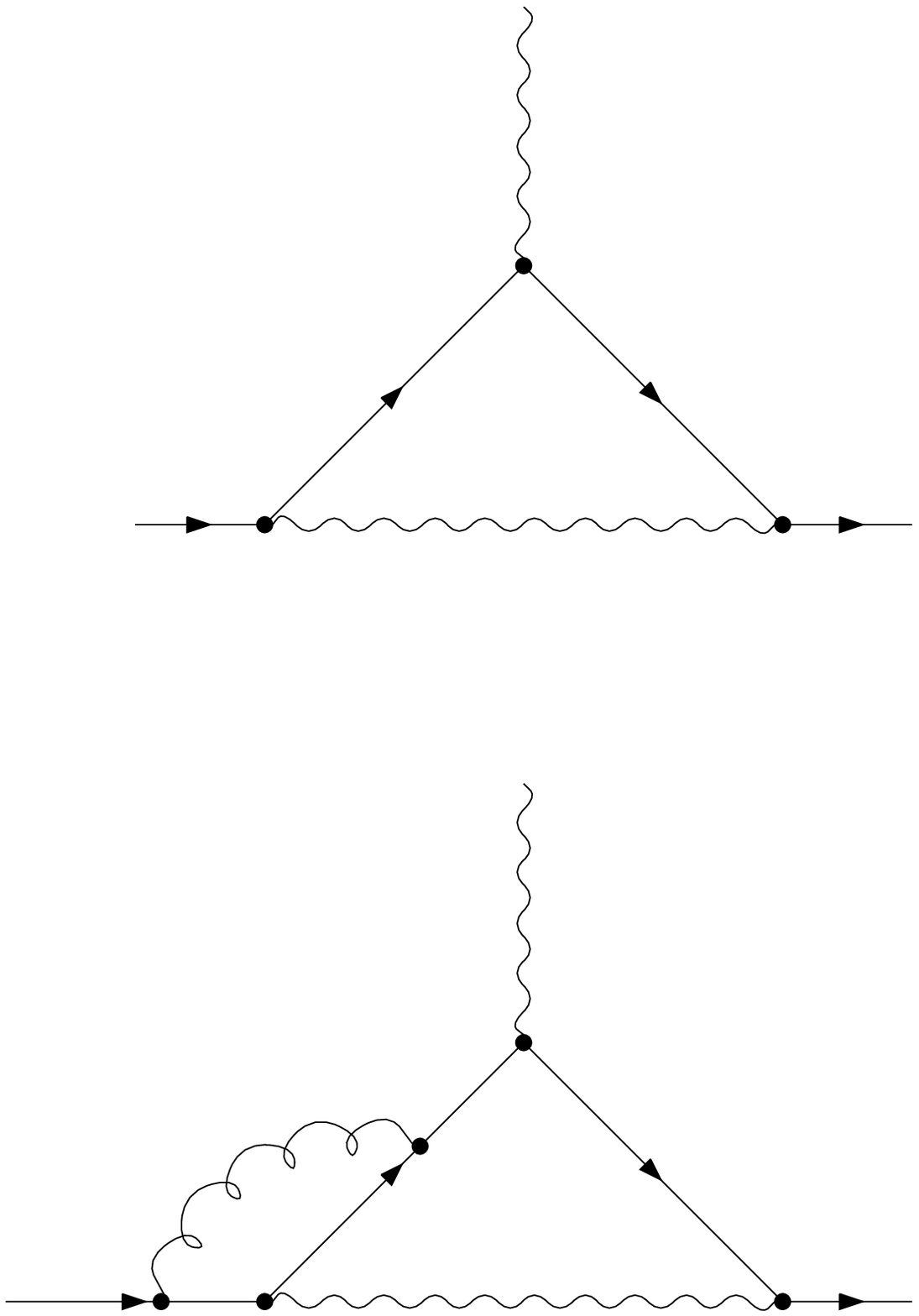}
\hspace{2cm}
\includegraphics[width=4cm,angle=0]{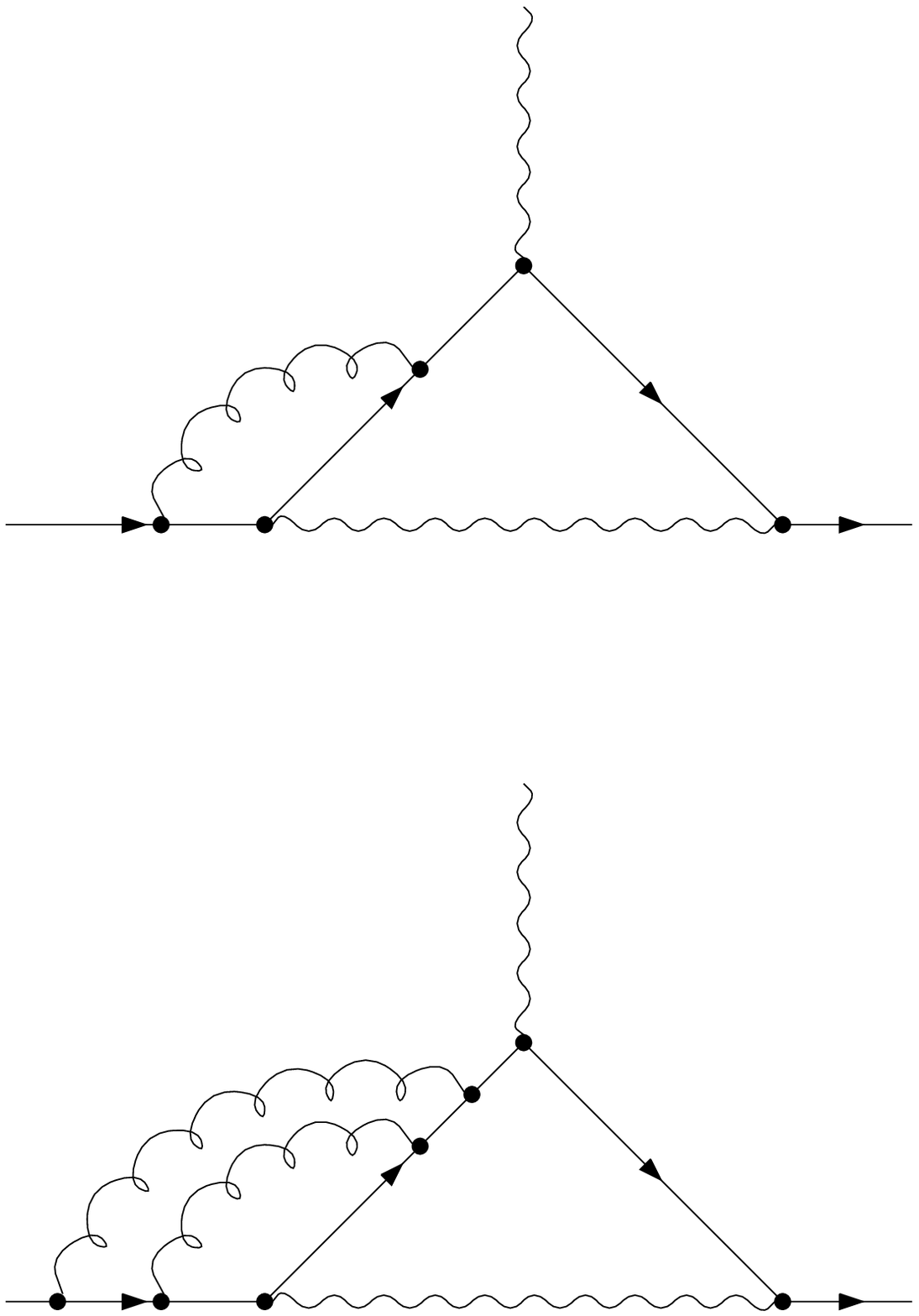}
\vspace{1cm}
\caption{Examples of Feynman diagrams contributing to $b \to s \gamma$
  at various orders in the renormalization-group-improved perturbation
  theory.}
\label{orders}
\end{center}
\vspace{-82mm} 
\hspace*{8mm} a. \hspace{55mm} b.\\
\hspace*{3cm} $\gamma$ \\[6mm]
\hspace*{15mm} $u,c,t$ \hspace{1cm} $u,c,t$\\[5.5mm]
\hspace*{14mm} $b$ \hspace{1cm} $W$\hspace{1cm} $s$
\hspace{20mm} $\als \ln \f{M_W^2}{m_b^2}$:~$\sim$~+50\% in amplitude\\[-1mm]
\hspace*{83mm} $\sim$~+100\% in BR\\[1mm]
\hspace*{8mm} c. \hspace{55mm} d.\\[2cm]
\hspace*{6mm} $\underbrace{\rm ~~~~~~~non-logarithmic 
                            \hspace{42mm} logarithmic~~~~~~}$\\
\hspace*{43mm} $\sim$~+20\% in BR\\[3mm]
\end{figure}

Most of this talk will be devoted to the loop-mediated $\bar{B} \to
X_s \gamma$ decay that is dominant in $\bar{B} \to X_{\rm no~charm}
\gamma$.  In order to make a theoretical prediction, we first need to
calculate the perturbative $b$-quark decay amplitudes to partonic
states $X_s^{(p)}$ and the photon. Later, the perturbative amplitudes
will enter directly into the expressions for hadronic branching
ratios.  A lot of effort has been devoted in the recent years to
calculating the $b$-quark decay amplitudes with better than 10\%
accuracy. Single gluon corrections (Fig.~\ref{orders}b) to the
one-loop $b\to s\gamma$ diagrams (Fig.~\ref{orders}a) increase the
predicted amplitude by around 50\%, and the branching ratio by around
100\%.  This effect is so large because the logarithm ~$\ln
\f{M_W^2}{m_b^2}$ is big and because the one-loop result is
accidentally quite small - it gives only about $\f{1}{5}$ of what is
naively expected. In order to achieve better than 10\% accuracy, one
needs to include the NLO QCD corrections (Figs.~\ref{orders}c and
~\ref{orders}d), i.e.  non-logarithmic parts of two-loop
diagrams\cite{2mtch,GHW96} and logarithmic parts of three-loop
diagrams.\cite{CMM97} The NLO corrections further increase the
predicted branching ratio by around 20\%. Both the LO and the NLO
calculations include resummation of large logarithms $\ln
\f{M_W^2}{m_b^2}$ from all orders of the perturbation series.

Resummation of large logarithms as well as further calculation of
hadronic decay rates is most conveniently performed in the framework
of an effective theory obtained from the SM by decoupling the heavy
electroweak bosons and the top quark. The Lagrangian of the effective
theory reads
\be 
{\cal L} =  {\cal L}_{QCD \times QED}(u,d,s,c,b) 
+ \f{4 G_F}{\sqrt{2}} V_{ts}^* V_{tb} \sum_{i=1}^8 C_i(\mu) O_i,
\ee
where the first term is just the QCD$\times$QED Lagrangian for the
light quarks, and the second term contains flavour-changing local
interactions $O_i$ of either 4 quarks or 2 quarks and gauge bosons. 
\be 
O_i = \left\{ \begin{array}{lll}
(\bar{s} \Gamma_i c)(\bar{c} \Gamma'_{\underline{i}} b), 
& i=1,2, & |C_i(\mu_b)| \sim 1,\\[2mm]
(\bar{s} \Gamma_i b) \sum_q (\bar{q} \Gamma'_{\underline{i}} q), 
& i=3,4,5,6,~~~~~~~~~ & |C_i(\mu_b)| < 0.07,\\[2mm]
\f{e m_b}{16 \pi^2} \bar{s}_L \sigma^{\mu \nu} b_R F_{\mu \nu}, 
& i=7, & |C_7(\mu_b)| \sim 0.3,\\[2mm]
\f{g m_b}{16 \pi^2} \bar{s}_L \sigma^{\mu \nu} T^a b_R G^a_{\mu \nu},~~
& i=8, & |C_8(\mu_b)| \sim 0.15.
\end{array} \right.  
\ee
The symbols $\Gamma_i$ and $\Gamma'_i$ in $O_1$,..., $O_6$ stand for
various products of the Dirac and colour matrices.  The
$\overline{MS}$-renormalized couplings $C_i$ at the scale $\mu_b \sim
m_b$ are known nowadays up to (and including) the following terms in
their perturbative expansion:\cite{CMM97,BM99}
\be
C_i(\mu_b) =   C_i^{(0)}(\mu_b) 
            + \f{\alpha_{em}}{\alpha_s(\mu_b)} C_i^{(0)em}(\mu_b)
            + \f{\alpha_s(\mu_b)}{4\pi}   C_i^{(1)}(\mu_b) + ...
\ee
Once the Wilson coefficients $C_i(\mu_b)$ are known, the $b$-quark
decay amplitudes are given by Feynman diagrams with single insertions
of the flavour-changing interactions, i.e. by matrix elements of the
operators $O_i$ between the appropriate partonic states.

\begin{figure}[htb]
\hspace*{4cm} $BR[\bar{B} \to \bar{K}^* \gamma] \times 10^5$\\[-8.5mm]
\hspace*{-3cm}
\begin{center}
\includegraphics[width=8cm,angle=0]{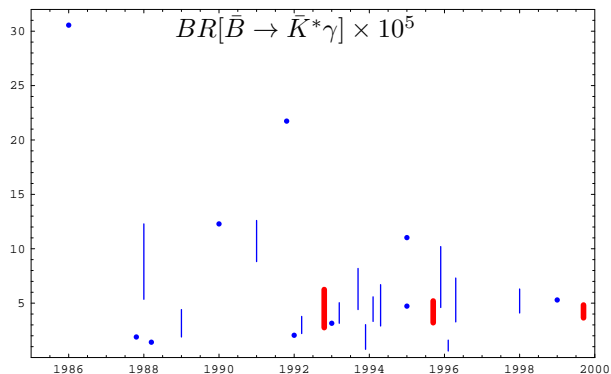}\\[-3mm]
\caption{Brief history of published predictions and experimental results 
for $BR[\bar{B} \to \bar{K}^* \gamma$].}
\ \\[-7mm]
\label{history}
\end{center}
\end{figure}

For the exclusive decay $\bar{B} \to \bar{K}^* \gamma$, one needs to
know matrix elements of those operators between the relevant hadronic
states: $\langle \bar{K}^* \gamma | O_i | \bar{B} \rangle$.  There
have been many attempts to calculate these matrix elements using quark
models, QCD sum rules, lattice and heavy quark symmetries (see
e.g.\cite{excl}). The history of published predictions is briefly
summarized in Fig.~\ref{history}. The (blue) thin bars and dots are
the theoretical predictions, while the (red) thick bars denote the
CLEO measurements.  Many recent theoretical papers on $\bar{B} \to
\bar{K}^* \gamma$ are not included in the plot because only
form-factors are discussed there, and no explicit number for the decay
rate is given.
A general conclusion one can derive from the $\bar{B} \to \bar{K}^*
\gamma$ studies is that they can help us in understanding
non-perturbative QCD, but $BR[\bar{B} \to \bar{K}^* \gamma]$ is not a
good place to look for new physics, given the present experimental and
theoretical results for the inclusive mode $\bar{B} \to X_s \gamma$.
However, observation of a sizable CP-asymmetry in the exclusive mode
would be a clear signal of new physics.\cite{AGS96}

For the inclusive decay $\bar{B} \to X_s \gamma$, the theoretical prediction
for the branching ratio can be made more precise by using the Operator
Product Expansion (OPE) within the Heavy Quark Effective Theory
(HQET).\cite{FLS94} We need to calculate
\be \label{square}
\sum_{X_s} \left| 
  C_7 \langle X_s \gamma | O_7 | \bar{B} \rangle
+ C_2 \langle X_s \gamma | O_2 | \bar{B} \rangle
+ ... \right|^2,
\ee
where dots stand for matrix elements of other operators that are
numerically less important. Let us first look at the "77" term, i.e.
the term proportional to $|C_7|^2$ in Eq.~(\ref{square}). This term
dominates in the perturbative calculation of the $b$-quark decay. In
full analogy to the semileptonic $B$-meson decay, we relate it via
optical theorem to the imaginary part of the elastic
forward scattering amplitude\\[2mm]
\hspace*{29mm}
\includegraphics[width=5cm,angle=0]{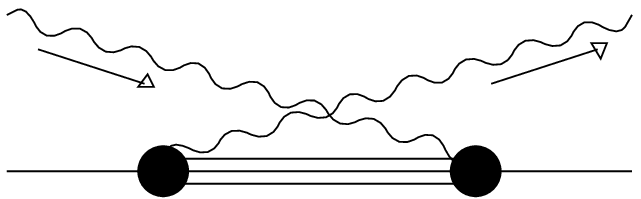}\\[-17mm]
\hspace*{4cm} $\gamma$ \hspace{2cm} $\gamma$\\[1mm]
\hspace*{35mm} $^q$ \hspace{32mm} $^q$\\
\hspace*{35mm} $\bar{B}$ \hspace{29mm} $\bar{B}$\\[-9mm]
\be
Im\{ \hspace{5cm} \} \equiv Im\;A
\ee
In this amplitude, we can perform OPE when the photon energies
$E_{\gamma}$ in the $\bar{B}$-meson rest frame are far from the
endpoint, i.e.  when $| m_B - 2 E_{\gamma} | >> \Lambda_{QCD}$.  Most
of the photons in $\bar{B} \to X_s \gamma$ have energies close to the
endpoint $E_{\gamma}^{\rm max} \simeq \f{1}{2}m_b$, so they do not
satisfy this requirement. Thus, at this first step, OPE gives us only
the tail of the photon spectrum.
\begin{figure}[h]
\hspace*{23mm} $Im\;E_{\gamma}$\\[-1cm]
\begin{center}
\includegraphics[width=8cm,angle=0]{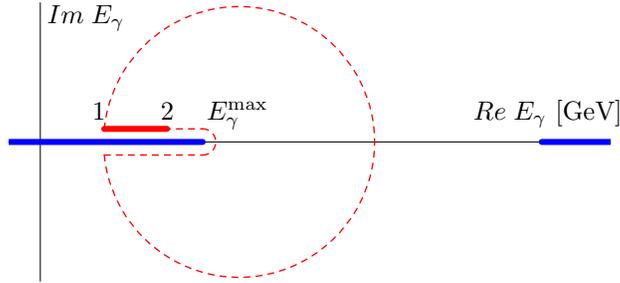}
\caption{Physical cuts in $A(E_{\gamma})$ (thick lines on the real axis) 
  and the relevant contour.}
\label{cuts}
\end{center}
\vspace*{-36mm}
\hspace*{29mm} 1 \hspace{5mm} 2 \hspace{2mm} $E_{\gamma}^{\rm max}$ 
\hspace{25mm} $Re\;E_{\gamma}$~[GeV]\\[25mm]
\end{figure}

Fortunately, we know analytic properties of the amplitude $A$ when
$E_{\gamma}$ is formally treated as complex.  We know that the
discontinuity of $A$ on the real axis is equal to $Im A$. Thus, if we
want to know the integral of $Im A$ from, say, 1~GeV to the endpoint,
we can find it by performing the integral of $A$ around the big circle
in Fig.~\ref{cuts}, where the condition for OPE is always fulfilled.
\be
\int_{1~{\rm GeV}}^{E_{\gamma}^{\rm max}} dE_{\gamma}\; E_{\gamma}^n \; Im \; A(E_{\gamma})
\sim \oint_{\rm big\;circle} dE_{\gamma}\; E_{\gamma}^n \; A(E_{\gamma}),
~~~~~n=0,1,2,...
\ee
We should better not go with $E_{\gamma}$ much below 1~GeV. There is
no problem in doing so for the "77" term, but there are problems with
other operators.\cite{LRW97,KLP95} Anyway, the region below 1~GeV is
hardly accessible experimentally, because of the $b\to c$ background.

The conclusion at this point is that we can predict the photon
spectrum for not too small and not too big energies, and moments of
the photon spectrum from not too small energies to the endpoint.
Making such predictions requires calculating matrix elements of
various local operators among $\bar{B}$-meson states. Matrix elements
of higher-dimensional operators are suppressed by higher powers of
$\Lambda/m_B$. Therefore, we can write a double expansion for the "77"
term, i.e. we can write
\bea 
\sum_{X_s} BR[ \bar{B} \to X_s \gamma]_{_{E_{\gamma} > 1~{\rm GeV}}}
&=& \left[ a_{00} + a_{02} \left(\f{\Lambda}{m_B}\right)^2 + ... \right]
\nonumber \\ && \hspace{-3cm}
+\f{\alpha_s(m_b)}{\pi} \left[  a_{10} + a_{12} \left(\f{\Lambda}{m_B}\right)^2 + ...\right]
+ {\cal O}\left[ \left(\f{\alpha_s(m_b)}{\pi}\right)^2 \right]
\nonumber \\ \label{double} && \hspace{-3cm}
+\mbox{ [ Contributions other than the "77" term].}
\eea
Here, the two terms not suppressed by $\Lambda/m_B$ are simply those
already found in the perturbative calculation of the $b$-quark decay.
The term proportional to $a_{02}$ turns out to give only an around ~$-3\%$
contribution.\cite{FLS94} The remaining terms in the first two lines
of Eq.~(\ref{double}) have stronger suppression factors, which makes
them negligible.

However, we need to ask whether a similar expansion can be written for
the third line of Eq.~(\ref{double}), i.e. for the contributions other
than the "77" term.  The answer to this question is {\em no}. These
remaining contributions contain, for instance, the huge effect from
intermediate $\psi$ that has been mentioned in the beginning of this
talk. The intermediate $\psi$ becomes important either due to
non-perturbative effects or because of big contributions at high
orders of perturbation theory that need to be resummed. Most probably,
both mechanisms are at work.

The intermediate $\psi$ contribution can be just subtracted from both
the experimental data and the theoretical predictions, using the
narrow peak approximation as in Eq.~(\ref{first}).  Other narrow
$c\bar{c}$ resonances hardly ever decay radiatively to charmless
states, so similar contributions from them are negligible.

Suppose we subtract the intermediate $\psi$ contribution.  Does the
sum of the remaining contributions take the form of a power series as
in the first two lines of Eq.~(\ref{double}), with the perturbatively
calculable leading term? It does not.\cite{LRW97,LLMW99} However, it
is hard to identify any obvious source of a big non-perturbative
effect in it. Operators containing no charm quark are suppressed by
their small Wilson coefficients. As far as the operators containing
the charm quark are concerned, we know that their contribution at the
leading order in $\als$ can be expressed as a power series\\[-1mm]

\hspace*{1mm}
\includegraphics[width=3cm,angle=0]{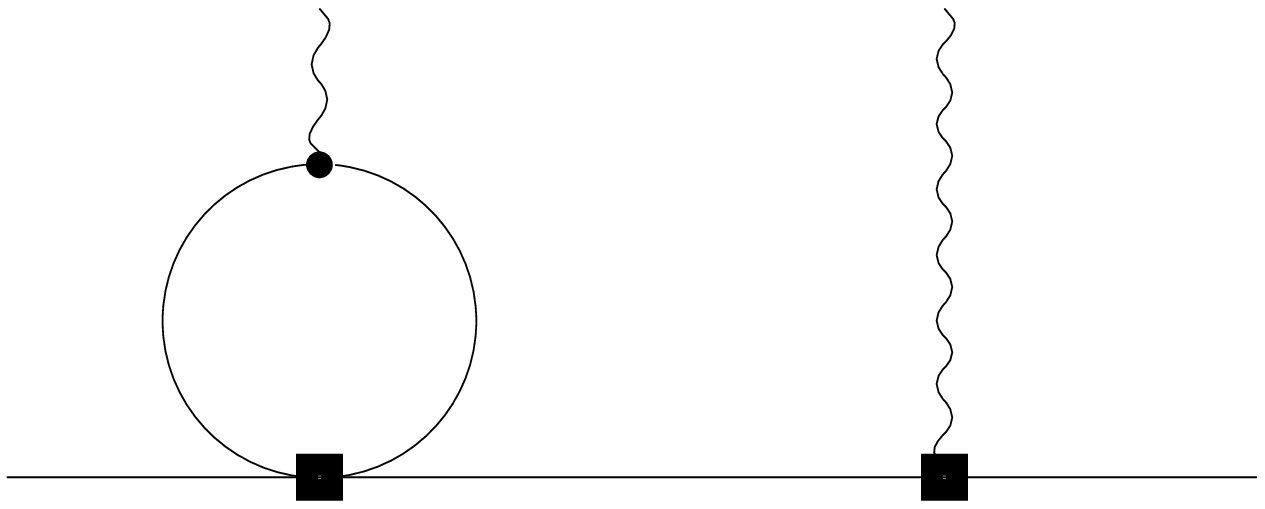}\\[-9mm]
\hspace*{9mm} $c$ \hspace{6.5mm} $c$\\[3mm]
\hspace*{14mm} $O_2$ \hspace{8mm} $O_7$\\[-16mm]
\be
\langle \bar{B} | \hspace{33mm} | \bar{B} \rangle = 
\mbox{(perturbative 0)}
+ \f{\Lambda^2}{m_c^2} \sum_{n=0}^{\infty} b_n 
         \left( \f{m_b \Lambda}{m_c^2} \right)^n,
\ee
which can be truncated to the leading $n=0$ term, because the
coefficients $b_n$ decrease fast with $n$. The
calculable\cite{LRW97,mc} $n=0$ term makes $BR[ \bar{B} \to X_s
\gamma]$ increase by around 3\%.

However, an analysis of non-perturbative effects in the matrix
elements of $O_1$ and $O_2$ at ${\cal O}(\alpha_s)$ is missing.
For instance,\\[-1mm]

\hspace*{13mm}
\includegraphics[width=3cm,angle=0]{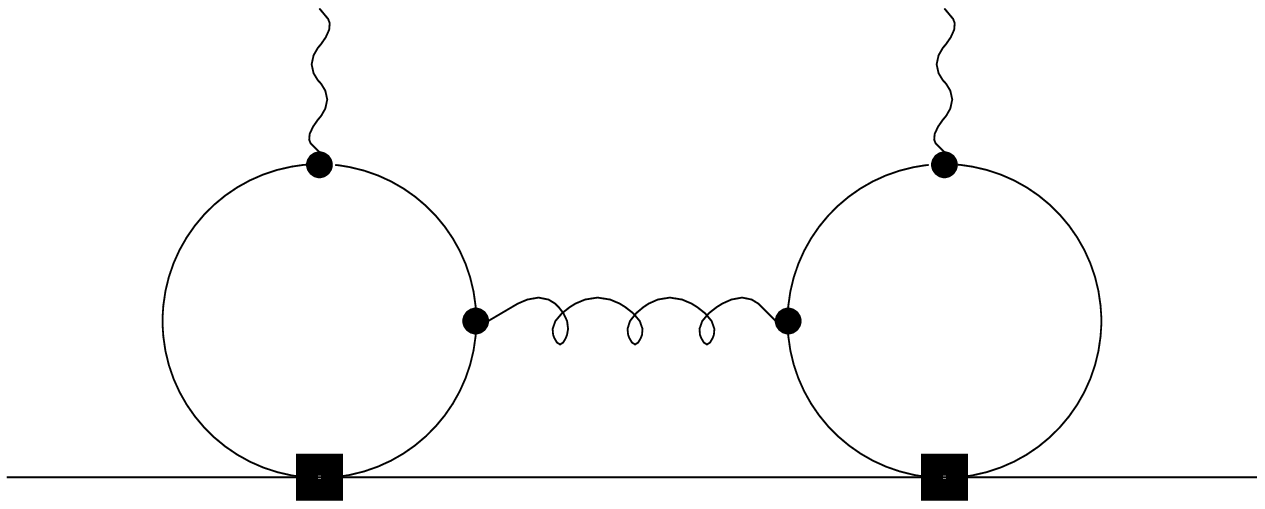}\\[-1cm]
\hspace*{31mm} {\rm hard}\\[4mm]
\hspace*{26mm} $O_2$ \hspace{8mm} $O_2$\\[-14mm]
\be
\langle \bar{B} | \hspace{33mm} | \bar{B} \rangle 
= A_{{\rm one}\mbox{-}{\rm loop}} + B_{\psi} + C_{\rm unknown},\\[2.5mm]
\ee
where $A_{{\rm one}\mbox{-}{\rm loop}}$ stands for the very small
(less than 1\% in BR) perturbative contribution from the gluon
bremsstrahlung at one loop, $B_{\psi}$ is a part of the (huge)
intermediate $\psi$ contribution, and $C_{\rm unknown}$ denotes the
remaining non-perturbative terms. Those remaining terms would not be
numerically important if they were either suppressed by
$\Lambda/m_{c,b}$, or small for other reasons, or could be absorbed
into the intermediate $\psi$ contribution. Unfortunately, I am not
aware of any sufficiently precise argument that any of these three
possibilities is realized.

In the following, I shall assume that one of these three possibilities
is realized.  In such a case, the hadronic decay rate is indeed
well-approximated by the partonic decay rate, up to small
non-perturbative corrections
\bea
\f{ \Gamma[ \bar{B} \to X_s \gamma]^{{\rm subtracted}\;\psi}_{E_{\gamma} > E_{\rm cut}}}
{ \Gamma[ \bar{B} \to X_c e \bar{\nu}_e]} \simeq
\f{ \Gamma[ b \to X_s \gamma]^{\rm perturbative~NLO}_{E_{\gamma} > E_{\rm cut}}}
{ \Gamma[ b \to X_c e \bar{\nu}_e]^{\rm perturbative~NLO}} \times
\nonumber \\ \label{bratio}
\times \left[ 1 + ({\cal O}(\Lambda^2/m_b^2)\simeq 1\%)
                + ({\cal O}(\Lambda^2/m_c^2)\simeq 3\%) \right].
\eea
The normalization to the semileptonic rate has been used here to
cancel uncertainties due to $m_b^5$, CKM-angles and some of the
non-perturbative corrections. One has to remember that
Eq.~(\ref{bratio}) becomes a bad approximation for $E_{\gamma}^{\rm
  cut} << 1$~GeV, and that non-perturbative corrections grow
dramatically\cite{KN99} when $E_{\gamma}^{\rm cut} > 2$~GeV.

For $E_{\rm cut} = 1$~GeV, Eq.~(\ref{bratio}) gives
\be \label{nbratio}
BR[ \bar{B} \to X_s \gamma]^{{\rm subtracted}\;\psi}_{E_{\gamma} > E_{\rm cut}}
= (3.29 \pm 0.33) \times 10^{-4},
\ee
where the dominant uncertainties originate from the uncalculated
${\cal O}(\alpha_s^2)$ effects and from the ratio $m_c/m_b$ in the
semileptonic decay (around 7\% each).

\begin{figure}[htb]
\begin{center}
\includegraphics[width=8cm,angle=0]{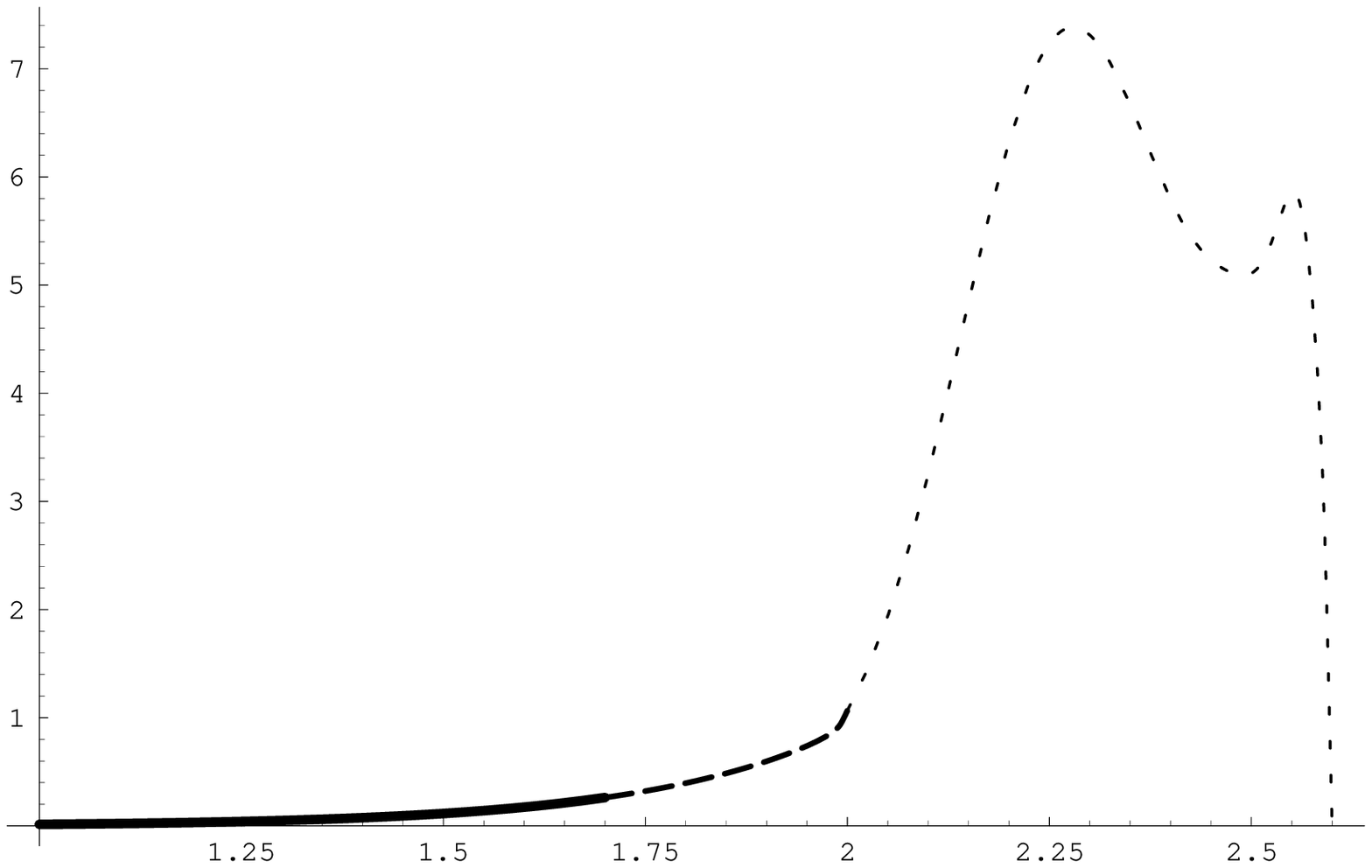}\\[-35mm]
\hspace*{-19mm}
$\f{d}{dE_{\gamma}} BR[\bar{B} \to X_s \gamma]^{{\rm subtracted}\;\psi} 
\times 10^4$\\[4cm]
\hspace*{95mm} $E_{\gamma}$~[GeV]\\[3mm]
\caption{An "artist view" of 
$\f{d}{dE_{\gamma}} BR[\bar{B} \to X_s \gamma]^{{\rm subtracted}\;\psi}$.}
\label{spectrum}
\end{center}
\end{figure}

Unfortunately, $E_{\rm cut} = 1$~GeV is not accessible experimentally
at present.  We need some prediction for the photon spectrum. The
solid line in Fig.~\ref{spectrum} describes the photon
spectrum\cite{LLMW99} in the region where the theoretical HQET
prediction is solid.  For larger energies, the less solid the line
becomes, the less solid the prediction is. In the peak region, it is
simply an "artist view" of how the spectrum could look like. However,
its normalization is fixed by Eq.~(\ref{nbratio}), and the size of the
visible $\bar{K}^*(892)$ peak is adjusted to the value measured by
CLEO.\cite{CLEO99b}

\ \\[-13mm]

If we knew the shape function of the $\bar{B}$ meson exactly, we could
make a solid prediction for the photon spectrum in the peak region,
too.\cite{KN99} Unfortunately, only models for the shape function are
available at present. Therefore, an optimal scenario for a comparison
between theory and experiment would be measuring the photon spectrum
above $\sim 2$~GeV without relying on any theoretical prediction for
its shape, and then extrapolating in a simple manner to the predicted
spectrum below $\sim 2$~GeV. The measured spectrum above 2~GeV would
provide\cite{N94} important information for extracting $V_{ub}$ from
$\bar{B} \to X_u l \nu$.

\newpage
The decay $\bar{B} \to X_d \; \gamma$ is theoretically more difficult
than $\bar{B} \to X_s \; \gamma$, because diagrams with up-quark loops
are no longer CKM-suppressed with respect to the remaining ones.
Thus, the theoretical accuracy is at best $\pm 30\%$, even for fixed
Wolfenstein parameters $\rho$ and $\eta$.

When the non-perturbative effects in up-quark loops are assumed to be
small, one obtains\cite{AAG98}
\bea 
\f{1}{2} \{ BR[ \bar{B} \to X_d \;\gamma] + BR[ B \to \bar{X}_d \;\gamma] \}
&\simeq& 2.43 [ (1-\bar{\rho})^2 + \bar{\eta}^2 - 0.35 (1-\bar{\rho}) 
\nonumber \\ && \hspace{-6cm}
+ 0.07] \times 10^{-5} = 1.61 \times 10^{-5}
~~(\mbox{for~} \bar{\rho}=0.11 \mbox{~~and~}  \bar{\eta}=0.32),
\eea
while the direct CP-asymmetries range from 7\% to 35\%. Estimates for the
exclusive channels are: %hep-ph/9911308
$BR[ B \to \rho^{\pm} \gamma] \in [ 1, 4] \times 10^{-6}$
~and~ 
$BR[ B \to (\rho^0, \omega) \gamma] \in [ 0.5, 2] \times 10^{-6}$.

The present experimental results for $\bar{B} \to X_s \gamma$ already
place severe constraints on extensions of the SM, like 2HDM, MSSM,
LR-models etc. Theoretical predictions for exotic contributions have
been recently calculated at NLO in many extensions of the
SM.\cite{beyond} These NLO effects are important only when the exotic
effects are large, but tend to cancel among each other and/or the SM
contribution, so that the present experimental constraints are
satisfied.

The CP-asymmetry in $B \to K^* \gamma$ is very small in the SM, but
could be significantly enhanced\cite{AGS96} in such extensions of the
SM, in which the flavour-changing interactions of the right-handed
$s$-quark are not suppressed by $m_s/M_W$, e.g. in the left-right
symmetric models. The present CLEO bound\cite{CLEO99b} on the
CP-asymmetry places important constraints on such models. Interesting
information on physics beyond the SM can be obtained from the
CP-asymmetry in the inclusive $\bar{B} \to X_s \gamma$ mode,
too.\cite{KN98}

\ \\
To conclude:

\begin{itemize}
  
\item{} The present theoretical prediction for $BR[\bar{B} \to X_s
  \gamma]$ in the SM agrees very well with the measurements of CLEO and
  ALEPH.  However, an analysis of non-perturbative effects at order
  ${\cal O}(\alpha_s)$ is necessary in order to make sure that the
  theoretical uncertainties are indeed around 10\%.
  
\item{} Future measurements of $BR[\bar{B} \to X_s \gamma]$ should
  rely as little as possible on theoretical predictions for the
  precise shape of the photon spectrum above $\sim 2$~GeV.

\end{itemize}

\noindent
Acknowledgements: I would like to thank P.~Ball, M.~Beneke,
G.~Buchalla, A.~Hoang, M.~Neubert and T.~Skwarnicki for helpful
discussions, and the organizers of the BCP3 conference in Taipei for
the invitation and warm hospitality.

\newpage
\setlength {\baselineskip}{0.2in}
 
\end{document}